\documentclass[12pt,preprint]{aastex}

\shortauthors{Harvey et al.}

\begin{document}

\slugcomment{Accepted for publication in the Astronomical Journal}

\title{Limits on Radio Continuum Emission from a Sample of Candidate
Contracting Starless Cores}

\author{Daniel W.A.\ Harvey,
        David J.\ Wilner, 
        James Di Francesco\altaffilmark{1}, \\
        Chang Won Lee\altaffilmark{2}, 
        Philip C.\ Myers}
\affil{Harvard-Smithsonian Center for Astrophysics, 60 Garden Street,
Cambridge, MA 02138}
\email{dharvey, dwilner, jdifran, cwlee, pmyers@cfa.harvard.edu}

\altaffiltext{1}{
currently at the Radio Astronomy Laboratory, 
University of California, Berkeley, Berkeley, CA 94720-3411;
e-mail: {\tt jdifran@astron.berkeley.edu}}

\altaffiltext{2}{
currently at the Taeduk Radio Astronomy Observatory,
Korea Astronomy Observatory, 36-1 Hwaam-dong, Yusung-gu, Taejon 305-348,
Korea; e-mail: {\tt cwl@hae.trao.re.kr}}

\and

\author{Jonathan P.\ Williams}
\affil{Department of Astronomy, University of Florida, Gainesville, FL 32611}
\email{williams@astro.ufl.edu}

\begin{abstract}
We used the NRAO Very Large Array to search for 3.6~cm continuum emission
from embedded protostars in a sample of 8 nearby ``starless'' cores that   
show spectroscopic evidence for infalling motions in molecular emission
lines. We detect a total of 13 compact sources in the eight observed fields 
to 5~$\sigma$ limiting flux levels of typically 0.09~mJy. None of these 
sources lie within $1'$ of the central positions of the cores, and they are
all likely background objects. Based on an extrapolation of the empirical
correlation between the bolometric luminosity and 3.6~cm luminosity for the
youngest protostars, these null-detections place upper limits of 
$\sim 0.1$~L${_\odot}(d/{\rm 140~pc})^{2}$ on the luminosities of
protostellar sources embedded within these cores. These limits, together
with the extended nature of the inward motions inferred from molecular line
mapping (Lee et al.\ 2001), are inconsistent with the inside-out collapse
model of singular isothermal spheres and suggest a less centrally condensed
phase of core evolution during the earliest stages of star formation.

\end{abstract}


\section{Introduction}

Star formation occurs within molecular clouds, behind large 
column densities of obscuring dust that hinders observations 
of its earliest evolutionary stages. The ``Class 0'' objects,
which are not detected at wavelengths shorter than the submillimeter,
represent the youngest protostars observed, with inferred ages of 
$\sim 10^4$~years (see Andr\'{e}, Ward-Thompson \& Barsony 1993, 
Andr\'{e} \& Montmerle 1994). Even younger protostars need to be 
identified to improve our understanding of the first stages of
protostellar collapse.

The ``starless'' dense cores, density enhancements within molecular
clouds that show no evidence for embedded protostars, may represent an
evolutionary stage of the star formation process prior to that of the
Class~0 objects. Lee et al.\ (1999) recently surveyed more than 200 nearby
starless cores in spectral lines of dense gas tracers. Seventeen cores were
identified with evidence for infalling motions, as suggested by the presence
of redshifted self-absorption in line profiles. These rare infall candidates 
provide an excellent chance to locate the very youngest protostars, whose 
low luminosity and reddened spectral energy distributions might have escaped 
detection at far-infrared wavelengths by the Infrared Astronomy Satellite 
(IRAS). 

A possible method to identify extremely young protostellar objects makes use
of the well-documented occurrence of simultaneous mass accretion (infall)
with mass expulsion (outflow) during star formation. From a sample of 29
Class~0 and Class~I protostellar sources, Anglada (1995) noted a strong
correlation between 3.6~cm radio continuum luminosity and bolometric
luminosity. The radio continuum emission arises from the shock-ionized inner
regions of a collimated outflow, while the bolometric luminosity arises
predominately from accretion (e.g.\ Rodriguez et al.\ 1982, Lada 1985). If
outflow is inextricably linked to accretion, the presence of a very young
protostar within a so-called ``starless core'' with evidence for infalling
motions can be deduced from the detection of compact radio continuum
emission. Using a variant of this concept, Visser (2000) found evidence for
three candidate protostars in a sample of 40 Lynds starless cores by
detecting high velocity ${}^{12}$CO(2--1) emission from previously unknown
outflows.

We present here observations of eight starless cores made with the 
Very Large Array (VLA) of the National Radio Astronomy Observatory\footnote{
The National Radio Astronomy Observatory is operated by Associated 
Universities Inc., under contract with the National Science Foundation.}   
to search for indications of embedded protostars through the detection 
of compact radio continuum emission from their nascent outflows.
The eight cores observed are the nearest
objects ($<250$~pc) from the survey of Lee et al.\ (1999) that show
redshifted self-absorption in at least two molecular spectral lines.

\section{Observations and Data Reduction}

The VLA observations of starless cores were made on 1999 March 6 and 1999
April 27 in the D configuration at 3.6~cm (8.46~GHz). The use of the VLA   
in its most compact configuration and at its most sensitive wavelength
maximizes the instrument's utility for detecting faint sources of radio
continuum emission. Table~1 lists the
target cores, calibrators, synthesized beam sizes, and the rms noise
levels achieved in the images. The central pointing positions for the cores
were taken from Lee et al.\ (1999). The uncertainties in these positions
with respect to the column density peaks are at the $\sim 15''$ level. For
the March observations of 4 cores at 4--5~hours Right Ascension (in the
Taurus region), 3C~48 was used as the flux calibrator (3.15~Jy at 3.6~cm).
For the April observations of 4 cores at 15--20~hours R.A.\, 3C~286 was used
as the flux calibrator (5.20~Jy at 3.6~cm). The flux calibration is expected
to be accurate to better than 10\%. Standard procedures in the AIPS software
package were used for all data calibration and imaging.

For each field, sources were located within the 5\farcm3 primary beam half
power diameter using the AIPS routine {\em Search and Destroy}, with a flux
threshold of $5 \times$ the rms noise in the cleaned image. Table~2 lists the
positions and fluxes of the 13 detected sources from the 8 fields. These
sources generally fall within the regions of extended molecular line emission,
but none are located within $1'$ of the adopted positions for the column
density peaks. This choice of criterion for positional coincidence is
arbitrary, but it is likely that all the detected sources are background
objects. Nevertheless, more millimeter/sub-millimeter observations must be
done to locate the column density peaks with better accuracy, to substantiate
the significance of the observed displacements. It remains conceivable that
one or more of the 3.6~cm sources may be tracing smaller clumps of material
embedded within the cores; this possibility could be tested by future higher
resolution observations. The observed source counts are consistent with
expectations for a background population. Based on a detection threshold of
$0.09$~mJy, the source properties from Condon (1984), and the calculation in
the appendix of Anglada et al.\ (1998), on average 1.0 background sources are
expected per field, for a total of $8\pm3$ sources, which is consistent within
2~$\sigma$ of the detected number. For the strongest source, L1521F~(1), a
reliable match is found in the NRAO VLA Sky Survey (NVSS) point source
catalog. The NVSS flux measurement at 21~cm (1.4~GHz) implies a radio spectral
index for this source of $-0.70\pm0.02$, which suggests synchrotron emission
from a background radio galaxy. This source is also the only resolved member
of the sample (the resolution of the cleaned images is $\sim 10''$). The
remaining unresolved sources show no matches in the SIMBAD astronomical
database.

\section{Discussion}

\subsection{Bolometric Luminosity Limits}

The lack of positional coincidence between any of the detected 
radio continuum sources and the centers of the starless cores 
suggests that none of these sources are associated with the cores. 
Hence, the observations allow us to place upper limits on the 3.6~cm 
radio continuum luminosity of any compact sources present in the cores, and
thus, upper limits on their bolometric luminosities following Anglada (1995).
Figure~1 shows the radio continuum and bolometric luminosity data for the 29
sources compiled by Anglada, together with a least-squares fit. 
The fitted correlation
\[ 
\log \: [ \: S_{3.6 \, \mathrm{cm}} \: d^2 \; (\mathrm{mJy~kpc}^2) \: ]
\; = \; (-2.1 \pm 0.1) \: + \: (0.7 \pm 0.1) \: \log \: [L_{\mathrm{bol}}
\; (\mathrm{L}_{\odot}) \: ] 
\]
allows the observed trend to be extrapolated to lower luminosities,
the regime of our sensitive VLA observations. While there is no 
observational support for the validity of this extrapolation, there 
are no changes expected in the relevant mechanisms at lower luminosity
levels that would suggest a breakdown of the correlation.

For the 4 Taurus cores (L1521F, TMC2, TMC1, CB23) at a distance
of 140~pc (Elias 1978), the correlation and 5~$\sigma$ upper limits
on the radio continuum luminosity imply bolometric
luminosity limits of $L\lesssim0.1$~L${}_\odot$. 
For the three Ophiuchus cores (L183, L158, L234E--S) at a distance 
of 165~pc (Chini 1981), the implied limits are
$L\lesssim 0.2$~L${}_\odot$. 
For L694--2, at an assumed distance of 250~pc that derives from 
an association with the cloud complex harboring the B335 core 
(Tomita, Saito \& Ohtani 1978), the implied bolometric luminosity limit 
is $L\lesssim0.7$~L${}_\odot$. 
Figure~1 shows the upper limits for the 3.6~cm luminosities as arrows. 
The uncertainty in these limits due to the uncertainty in the fitted
correlation is $\sim 50\%$.

\subsection{Comparison with IRAS Limits}

In some circumstances, the limits on protostar luminosity implied 
by the VLA null-detections represent a substantial improvement on 
limits implied by the lack of an IRAS point source.
For objects like starless cores that may be characterized by 
low temperatures, the limit from the longest wavelength IRAS band 
(100~$\mu$m) most strongly constrains luminosity. For isolated objects
at high Galactic latitudes ($|b| > 50^{\circ}$), the IRAS Point Source
Catalog has a 50\%-completeness limit of 1.0~Jy at 100~$\mu$m (Beichman
et al.\ 1988). Nearer to the plane of the Galaxy, in more complicated
regions such as Ophiuchus or Orion, the limits are often much higher
due to extended ``Cirrus'' emission, and confusion with other sources.
For example, VLA~1623, the prototypical Class~0 object, escaped detection
by IRAS despite a fairly warm dust temperature ($T \simeq 20$~K) and
substantial bolometric luminosity $L \sim 0.5$--$2.5$~L${}_{\odot}$
(Andr\`{e}, Ward-Thompson \& Barsony 1993). For this core, crowding with
nearby sources causes the IRAS limiting flux to be raised to 45~Jy
(Ward-Thompson 1993).

To calculate appropriate IRAS limits on bolometric luminosity,
we assume that the spectral energy distributions of Class~0 objects and
dense cores may be modeled reasonably well by modified graybodies of the
form:
\[F_{\nu} = B_{\nu}(T_{dust}) (1-\exp[-\tau_{\nu}]) \Omega_{S} \ , \]
where $B_{\nu}(T_{dust})$ denotes the Planck function at frequency $\nu$ for
a dust temperature $T_{dust}$, $\tau_{\nu}$ is the dust optical depth, 
and $\Omega_{S}$ the solid angle subtended by the source (Gordon 1988,
Ward-Thompson, Andr\'{e} \& Kirk 2001). The optical depth is proportional
to the mass opacity of the dust, which is generally assumed to follow a
power law with frequency. The power law index is uncertain for the dust in
dense cores, but bounded to a small range (Ossenkopf \& Henning 1994). Here, 
we adopt $\tau_{\nu} \propto \nu^{1.5}$, as found for VLA~1623 by Andr\`{e}
et al.\ 1993. The solid angle subtended by the source is constrained to be 
$\lesssim 2^{\prime}$ in diameter, the resolution of IRAS at 100~$\mu$m.   
The maximum bolometric luminosity of such a graybody that may be 
hidden below a 1.0~Jy IRAS completeness limit is temperature dependent,
roughly 1~L${}_{\odot}$ for $T_{dust}=10$~K, and falling to
0.1~L${}_{\odot}$ for $T_{dust}=14$~K, for a source at Taurus distance that 
is optically thin for $\lambda \gtrsim 100$~$\mu$m. This result is
similar to the 0.05--0.1~L${}_{\odot}$ limit for Taurus obtained 
by Myers et al.\ (1987), although the analysis presented here 
is more applicable to Class~0 objects.

Since the cores in this study are often viewed superposed against a 
parent molecular cloud, the 100~$\mu$m point source completeness limit 
may be worse than the nominal value of 1~Jy for isolated objects. 
For a $\sim 3$~Jy limit, a source as bright as 0.3~L${}_{\odot}$ at
140~pc would have been undetected by IRAS even with a dust temperature 
of 14~K. The presence of even a small amount of
(visible) warm dust would strongly increase the emission at the shorter
wavelengths, and substantially lower the bolometric luminosity that could
be hidden from IRAS. But such warm dust would be located close to the
protostar, in the central regions of the core, in which case the solid
angle subtended by the warm component would be small, and it would be
shielded from view by absorption in the intervening layers of cold dust.
To illustrate this scenario, consider an example of a cold core (14~K)
that surrounds a warm dust component (40~K) that represents $\sim 1$\% of
the total dust mass. If this core has an optical thickness that is only
one-tenth that of VLA~1623, making it optically thick for
$\lambda \lesssim 20$~$\mu$m, then it could produce a bolometric luminosity
of 0.1~L${}_{\odot}(d/{\rm 140~pc})^{2}$ but remain undetected by IRAS with
a 100~$\mu$m detection limit of $\sim3$~Jy. In these circumstances, the
limits on protostar luminosity implied by the VLA null-detections are
at least comparable to, and in many cases significantly lower than the
corresponding IRAS limits.

\subsection{Implications for the Infall Process}

The null-detection of an embedded point source in a contracting core has
implications for the nature of the infall process. Tafalla et al.\ (1998)
presented a detailed discussion of these implications pertaining to L1544, a
starless core in Taurus with evidence for infalling motions and comparable
3.6~cm radio continuum limits (Williams et al.\ 1999).  In short, the low
luminosities and extended sizes of the observed regions of inward motions
are inconsistent with the widely applied model of inside-out collapse of a
singular isothermal sphere (Shu 1977). In this model, collapse leads to the
formation of a central point source with luminosity
$L \sim a^6 t / G R_{\ast}$, where $a$ is the effective sound speed, $t$ is
the time since the onset of collapse, and $R_{\ast}$ is the protostellar
radius (Shu, Adams, \& Lizano 1987). Taking $R_{\ast} \simeq 3 R_{\odot}$
(Stahler 1988) and $a\simeq 0.19$~km~s${}^{-1}$ ($T \simeq 10$~K), the model
predicts $L\sim 3\: (t/10^5\, \mathrm{yr})\:$~L${}_{\odot}$. Infall onto a
disk would reduce this luminosity (Kenyon et al.\ 1993), but these very
young objects undetected by IRAS show no evidence for disks, nor are large
disks predicted at such early times (Terebey, Shu \& Cassen 1984). In the
context of the Shu model, and using the Anglada
correlation between radio continuum luminosity and bolometric luminosity,
for an embedded source at Taurus distance to remain undetected by our VLA
observations would require $t \lesssim 4 \times 10^3$~years, corresponding
to an infall radius of only $R_{\mathrm{inf}} \lesssim 8 \times 10^{-4}$~pc.
For a source at Ophiuchus distance, the corresponding values are
$t \lesssim 7 \times 10^3$~years, and
$R_{\mathrm{inf}} \lesssim 1 \times 10^{-3}$~pc, while at the greater
distance of L694--2, the values are $t \lesssim 2 \times 10^5$~years, and
$R_{\mathrm{inf}} \lesssim 0.04$~pc. Lee et al.\ (2001) have
computed sizes for the infalling regions in six of the eight cores that
were observed in this study (TMC2, TMC1, L183, L158, L234E--S, and
L694--2) by analyzing the extent of redshifted self-absorption in lines of
CS and N$_2$H$^+$. The infall radii so derived range from 0.07 pc for L158
up to 0.17 pc for TMC1, and are typically ~0.1 pc. The method used to
calculate these radii is fairly simple, and more detailed modelling may
refine the exact values. Nevertheless, the observed infall radii are very
much larger than the limits on radii derived from the strict luminosity
constraints found here assuming the inside-out collapse scenario. Moreover,
the ambipolar diffusion process in a strongly sub-critical core apparently
cannot explain the high velocities of infalling material observed in these
cores (Lee et al.\ 2001; Ciolek \& Mouschovias 1995). The cores studied here
represent a sample whose inward motions are inconsistent with the standard
theories of low mass star formation.

\section{Summary}

We used the VLA to search for 3.6~cm radio continuum emission from
embedded outflows in eight ``starless cores'' that show evidence for
infalling motions in the profiles of molecular spectral lines (Lee et
al.\ 1999). We detected a total of 13 compact sources of radio continuum
emission. None of the sources are within an arcminute of the column
density peaks, and the source counts are consistent with expectations for
a background population. Based on these apparent null-detections and an
observed correlation between outflow luminosity and bolometric
luminosity from protostellar objects, we place approximate upper limits on
the luminosities of any embedded protostars (with uncertainty $\sim 50$\%)
of $\lesssim 0.1$~L${_\odot}(d/{\rm 140~pc})^{2}$. 
These limits, together with the extended nature of the inward motions
inferred from molecular line mapping (Lee et al.\ 2001), are inconsistent 
with the inside-out collapse model of singular isothermal spheres and 
suggest a less centrally condensed phase of core evolution during the
earliest stages of star formation.

\acknowledgements
This research has made use of the SIMBAD database, operated at CDS,
Strasbourg, France.

\clearpage

\begin{deluxetable}{lccccc}
\tablenum{1}
\tablewidth{0pt}
\tablecolumns{6}
\tablecaption{Dense Cores Observed at 3.6~cm}
\tablehead{ 	& \multicolumn{2}{c}{Phase Center} & & & \colhead{RMS} \\
	    	&  \multicolumn{2}{c}{\hrulefill} &
		& \colhead{Beam} &  \colhead{Noise}\\
\colhead{Source}& \colhead{$\alpha$ (J2000)} & \colhead{$\delta$ (J2000)} &
\colhead{Calibrator} & \colhead{($''\times''$)} & 
\colhead{($\mu$Jy)} }

\startdata
L183 \dotfill & $15^{h \, } 54^{m} 06\hbox{$.\!\!^{s}$}5$ & $-02^{\circ}
51^{\prime} 39^{\prime\prime}$ & $1543-079$ & $11.6 \times 
9.8$ & 15\\
L158 \dotfill & $16 \! \hfill 47 \hfill 23.2$ & $-13 \hfill
59 \hfill 21 \hfill$ & $1543-079$ & $13.\hfill5 \times
9.\hfill8$ & 16\\
L234E--S \ldots & $16 \! \hfill 48 \hfill 08.6$ & $-10 \hfill
57 \hfill 24 \hfill$ & $1707+018$ & $12.\hfill9 \times
9.\hfill9$ & 18\\
L694--2 \dotfill & $19 \! \hfill 41 \hfill 04.5$ & $+10 \hfill
57 \hfill 02 \hfill$ & $1950+081$ & $11.\hfill9 \times
9.\hfill1$ & 20\\
L1521F \dotfill & $04 \! \hfill 28 \hfill 39.8$ & $+26 \hfill
51 \hfill 35 \hfill$ & $0432+416$ & $10.\hfill3 \times
9.\hfill1$ & 22\\
TMC2 \dotfill & $04 \! \hfill 32 \hfill 48.7$ & $+24 \hfill 
24 \hfill 12 \hfill$ & $0432+416$ & $10.\hfill4 \times
9.\hfill1$ & 18\\
TMC1 \dotfill & $04 \! \hfill 41 \hfill 33.0$ & $+25 \hfill
44 \hfill 44 \hfill$ & $0432+416$ & $10.\hfill4 \times
9.\hfill1$ & 17\\ 
CB23 \dotfill & $04 \! \hfill 43 \hfill 31.5$ & $+29 \hfill 
39 \hfill 11 \hfill$ & $0432+416$ & $10.\hfill3 \times
9.\hfill1$ & 18\\ 
\enddata
\end{deluxetable}

\clearpage
\begin{deluxetable}{lcccc}
\tablenum{2}
\tablewidth{0pt}
\tablecolumns{5}
\tablecaption{Sources Detected at 3.6~cm}
\tablehead{ & \multicolumn{2}{c}{Position\tablenotemark{a}} & &
		\colhead{Angular} \\
	    & \multicolumn{2}{c}{\hrulefill} &
		\colhead{Flux\tablenotemark{b}} &
		\colhead{Displacement\tablenotemark{c}}\\
	    \colhead{Source} & \colhead{$\alpha$ (J2000)} &
		\colhead{$\delta$ (J2000)} & \colhead{(mJy)} &
		\colhead{($''$)}}
\startdata
\parbox{10mm}{L158} (1) \dotfill & $16^{h \, } 47^{m} 16\hbox{$.\!\!^{s}$}7$
& $-14^{\circ} 00^{\prime} 57^{\prime\prime}$ & \hfill 0.27 & $\hfill 135
\hfill$ \\
\parbox{10mm}{\hfill} (2) \dotfill & $16 \! \hfill 47 \hfill 24.0$ & $-14
\hfill 01 \hfill 15 \hfill$ & \hfill 0.37 & $\hfill 115 \hfill$\\
\parbox{16mm}{L234E--S} (1) \ldots & $16 \! \hfill 48 \hfill 05.0$ & $-10
\hfill 59 \hfill 20 \hfill$ & \hfill 0.55 & $\hfill 128 \hfill$\\
\parbox{16mm}{\hfill} (2) \ldots & $16 \! \hfill 48 \hfill 06.5$ & $-10
\hfill 56 \hfill 07 \hfill$ & \hfill 0.25 & $\hfill 83 \hfill$\\
\parbox{14mm}{L694--2} (1) \dotfill & $19 \! \hfill 40 \hfill 57.2$ & $+10
\hfill 57 \hfill 30 \hfill$ & \hfill 2.60 & $\hfill 111 \hfill$\\
\parbox{14mm}{L1521F} (1) \dotfill & $04 \! \hfill 28 \hfill 41.2$ & $+26
\hfill 53 \hfill 55 \hfill$ & \hfill 10.75 & $\hfill 142 \hfill$\\
\parbox{14mm}{TMC2} (1) \dotfill & $04 \! \hfill 32 \hfill 42.6$ & $+24
\hfill 23 \hfill 24 \hfill$ & \hfill 0.24 & $\hfill 96 \hfill$\\
\parbox{14mm}{\hfill} (2) \dotfill & $04 \! \hfill 32 \hfill 54.2$ & $+24
\hfill 22 \hfill 43 \hfill$ & \hfill 0.13 & $\hfill 116 \hfill$\\
\parbox{14mm}{TMC1} (1) \dotfill & $04 \! \hfill 41 \hfill 25.4$ & $+25
\hfill 42 \hfill 53 \hfill$ & \hfill 0.67 & $\hfill 151 \hfill$\\
\parbox{14mm}{\hfill} (2) \dotfill & $04 \! \hfill 41 \hfill 25.7$ & $+25
\hfill 43 \hfill 49 \hfill$ & \hfill 0.75 & $\hfill 113 \hfill$\\
\parbox{14mm}{\hfill} (3) \dotfill & $04 \! \hfill 41 \hfill 39.0$ & $+25
\hfill 43 \hfill 24 \hfill$ & \hfill 0.36 & $\hfill 114 \hfill$\\
\parbox{14mm}{CB23} (1) \dotfill & $04 \! \hfill 43 \hfill 34.9$ & $+29   
\hfill 38 \hfill 05 \hfill$ & \hfill 0.14 & $\hfill 79 \hfill$\\
\parbox{14mm}{\hfill} (2) \dotfill & $04 \! \hfill 43 \hfill 35.0$ & $+29 
\hfill 37 \hfill 13 \hfill$ & \hfill 0.26 & $\hfill 127 \hfill$\\
\enddata
\tablenotetext{a}{Position errors are $\sim 1''$}
\tablenotetext{b}{Corrected for primary beam attenuation}
\tablenotetext{c}{From Phase Center in Table~1}
\end{deluxetable}

\clearpage
\begin{figure}[htb]
\figurenum{1}
\setlength{\unitlength}{1in}
\begin{picture}(6,5.0)
\put(-0.5,-4.9){\includegraphics{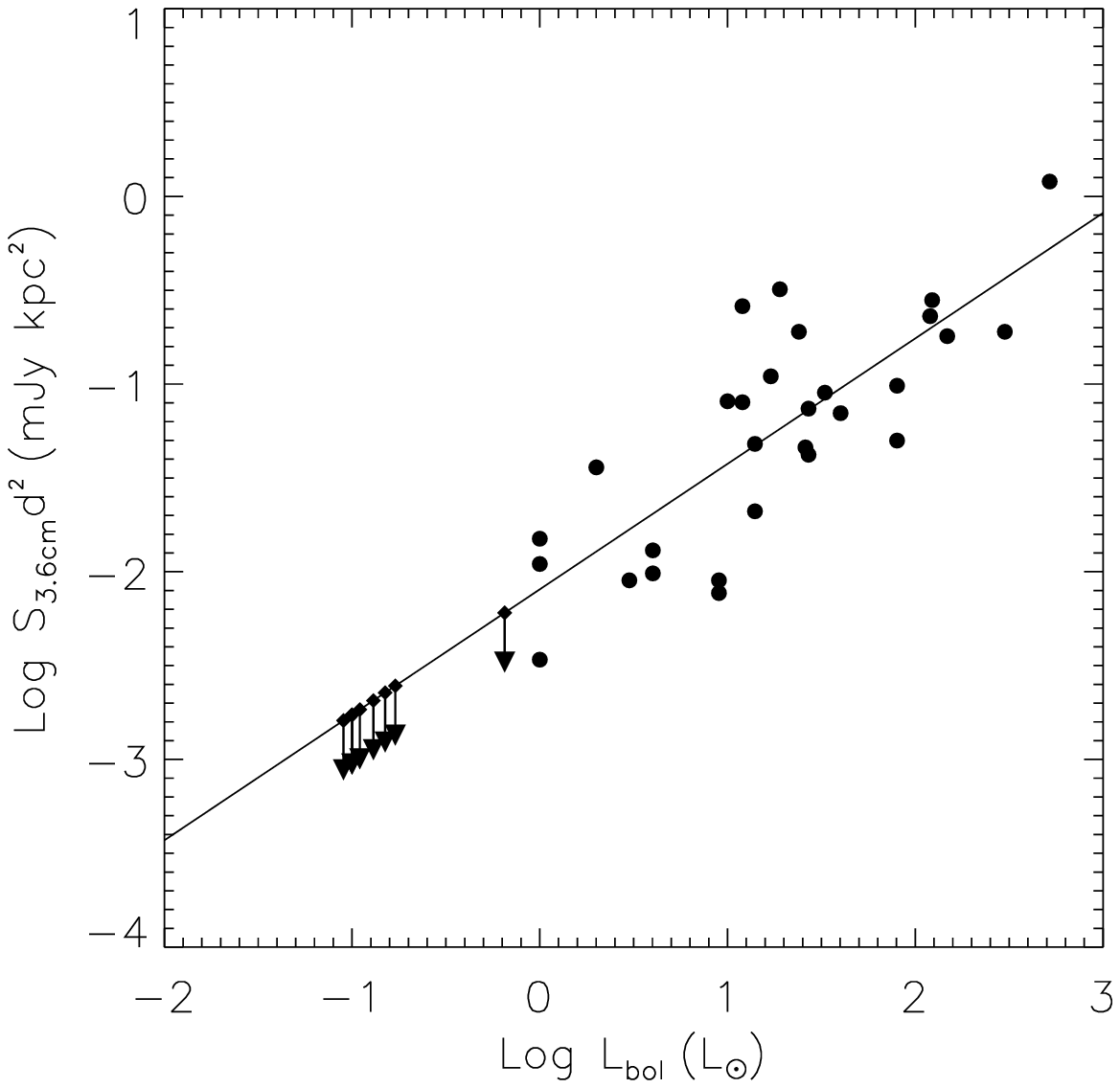}}
\end{picture}
\caption{Log-log plot of 3.6~cm luminosity against bolometric luminosity
for 29 outflow sources from Anglada (1995). A linear least-squares fit 
(r=0.79) indicated by the line has been used to extrapolate the observed 
trend to the sensitivity of our VLA observations. The fitted line has been 
used to calculate approximate upper limits on the bolometric luminosity of 
any protostar that might remain undetected in our observations. These upper
limits are marked with arrows.}
\end{figure}


\clearpage
\begin{references}

\reference{}
Andr\'{e}, P., \& Montmerle, T.\ 1994, \apj, 420, 837

\reference{}
Andr\'{e}, P., Ward-Thompson, D., \& Barsony, M.\ 1993, \apj, 406, 122

\reference{}
Anglada, G., Villuendas, E., Estalella, R., Beltr\'{a}n, M.T.,
Rodr\'{i}guez, L.F., Torrelles, J.M., \& Curiel, S.\ 1998, \aj, 116, 2953

\reference{}
Anglada, G.\ 1995, RMxAA, 1, 67

\reference{}
Beichman, C.\ et al.\ 1988, {\em Infrared Astronomical Satellite (IRAS)
Catalogs and Atlases, vol.\ 1, Explanatory Supplement}, NASA RP-1190
(Washington, DC: GPO)

\reference{}
Chini, R.\ 1981, \aa, 99, 346

\reference{}
Ciolek, G.E.\ \& Mouschovias, T.C\small{H}.\ 1996, \apj, 468, 749

\reference{}
Condon, J.J.\ 1984, \apj, 287, 461

\reference{}
Elias, J.H.\ 1978, \apj, 224, 857 

\reference{}
Gordon, M.A.\ 1988, \apj, 331, 509

\reference{}
Kenyon, S.J., Calvet, N.\ \& Hartmann, L.\ 1993, \apj, 414, 676

\reference{}
Lada, C.J.\ 1985, \araa, 23, 267

\reference{}
Lee, C.W., Myers, P.C., \& Tafalla, M.\ 2001, \apjs, 136, 703

\reference{}
Lee, C.W., Myers, P.C., \& Tafalla, M.\ 1999, \apj, 526, 788

\reference{}
Myers, P.C., Fuller, G.A., Mathieu, R.D., Beichman, C.A., Benson, P.J., 
Schild, R.E., \& Emerson, J.P.\ 1987, \apj, 319, 340

\reference{}
Ossenkopf, V.\ \& Henning, T.\ 1994, \aap, 291, 943

\reference{}
Rodriguez, L.F., Carral, P., Moran, J.M., \&  Ho, P.T.P.\ 1982, \apj, 260,
635

\reference{}
Shu, F.H.\ 1977, \apj, 214, 488

\reference{}
Shu, F.H., Adams, F.C., \& Lizano, S.\ 1987, \araa, 25, 23

\reference{}
Stahler, S.W.\ 1988, \apj, 332, 804

\reference{}
Tafalla, M., Mardones, D., Myers, P.C., Caselli, P., Bachiller, R., \&
Benson, P.J.\ 1998, \apj, 504, 900

\reference{}
Terebey, S., Shu, F.H.\ \& Cassen, P.\ 1984, \apj, 286, 529

\reference{}
Tomita, Y., Saito, T.\ \& Ohtani, H.\ 1979, \pasj, 31, 407

\reference{}
Visser, A.\ 2000, Ph.D.\ thesis, University of Cambridge (UK)

\reference{}
Ward-Thompson, D.\ 1993, \mnras, 265, 493

\reference{}
Ward-Thompson, D., Andre\'{e}, P.\ \& Kirk, J.M.\ 2001, \mnras, {\em in
press}

\reference{}
Williams, J.P., Myers, P.C., Wilner, D.J. \& Di Francesco, J. 1999,
\apj, 513, L61

\end{references}
\end{document}